\documentclass[preprint,12pt]{elsarticle}

\usepackage{amssymb}
\usepackage{amsfonts} 
\usepackage{amsmath}
\usepackage{amssymb}
\usepackage{braket}
\usepackage{breqn}
\usepackage{bm} 
\usepackage{dcolumn}

\usepackage{float}
\usepackage{graphicx}
\usepackage{lipsum}
\usepackage{mathtools}

\usepackage{verbatim}
\usepackage[utf8]{inputenc} 
\usepackage{soul,xcolor,colortbl}
\usepackage{hyperref}\hypersetup{colorlinks=true,citecolor=blue,linkcolor=blue,urlcolor=blue} 

\begin{document}

\begin{frontmatter}



\title{Characterization and visualization of grain boundary disconnections}


\address[add1]{Lawrence Livermore National Laboratory, Livermore, CA 94550, USA}

\author[add1]{I. S. Winter\footnote{Corresponding Author: winter24@llnl.gov}}            
\author[add1]{T. Oppelstrup}
\author[add1]{T. Frolov}
\author[add1]{R. E. Rudd}
            
\begin{abstract}
We introduce a method to visualize dislocations along grain boundaries at the atomic level. It uses an atomic-level Nye tensor, representing the dislocation density. To calculate the Nye tensor at grain boundaries, we extend the Hartley-Mishin strain gradient calculation to the displacement shift complete lattice. We show that the method is effective in visualizing disconnections and the dislocation content of grain boundary phase junctions in body-centered cubic tungsten, as well as face-centered cubic copper. In addition, we use the method to characterize the morphology of a two-dimensional grain boundary phase nucleus in a symmetric tilt grain boundary in tungsten.  This method can be applied to both bulk dislocations and grain boundary disconnections, which makes it ideal for studying the interactions and reactions of bulk dislocations with grain boundaries, and grain boundary disconnections.
\end{abstract}

\end{frontmatter}


\section{Introduction}
\label{sec:introduction}

Disconnections, or secondary grain boundary dislocations, are line defects that exist within grain boundaries. Disconnections are dislocations: they are characterized by a displacement discontinuity quantified by a Burgers vector \cite{Shober1970}, but the Burgers vector of a disconnection can be a fraction of that of a bulk dislocation. Over the past 50 years disconnections have been studied frequently, as disconnection motion can be tied to grain boundary migration and microstructural evolution \cite{Shober1970,HIRTH1973929,Sun1982,BABCOCK19892357,BABCOCK19892367,HIRTH19964749,HAN2018386}. Recently disconnection-based models of grain boundary (GB) migration \cite{Chen4533,CHEN2020412}, triple junction motion \cite{Thomas8756}, GB phase nucleation \cite{winter2021nucleation}, and grain growth stagnation \cite{Chen33077} have been developed. From these studies it is clear that the disconnection can be utilized as an effective building block for understanding microstructural evolution.

Large-scale molecular dynamics simulations have shown themselves to be capable of accurately modeling various aspects of plasticity \cite{Zepeda2017, wehrenberg2017grainrot, Zepeda2021}, and thus open the path towards large-scale atomistic simulations of microstructural evolution.  A key tool used to understand large-scale simulations is the dislocation extraction algorithm (DXA), which allows one to quantify and visualize dislocations within a crystal \cite{Stukowski_2010, Stukowski_2012}. However, the application of DXA to grain boundaries, and thus polycrystalline systems, is limited. This is due to the fact that DXA works by finding ``bad" regions within a system, regions that do not correspond to the perfect crystal, and running Burgers circuit calculations around these ``bad" regions. Stukowski et al.\ developed a work-around such that DXA can be applied to plane defects such as stacking faults and grain boundaries \cite{Stukowski_2012}, however this method requires for the algorithm to be ``trained" to recognize plane defects as regions of ``good" crystal.

It is with this constraint in mind that we introduce an alternative procedure to visualize dislocations in grain boundaries. Our approach is based on the work of Hartley and Mishin, who developed a methodology to estimate the Nye tensor \cite{NYE1953153} at an individual atom  \cite{HARTLEY20051313,HARTLEY200518}. This atomic Nye tensor gives the dislocation density for a given atom, and has become a well-utilized tool in visualizing bulk dislocations. We show that the method of Hartley and Mishin can be extended to treat dislocations within grain boundaries if one knows the relative orientations of the two grains. We show the effectiveness of this method by applying it to a disconnection in a  $\Sigma 5 (210)[001]$ tilt GB in body-centered cubic (BCC) W, a GB phase junction in a $\Sigma 17 (410)[001]$ tilt GB in BCC W, and a $\Sigma 5 (210)[001]$ tilt GB in face-centered cubic (FCC) Cu.

\section{Theory}
\label{sec:theory}

\begin{figure}[ht!]
\centering
\includegraphics[width=\textwidth]{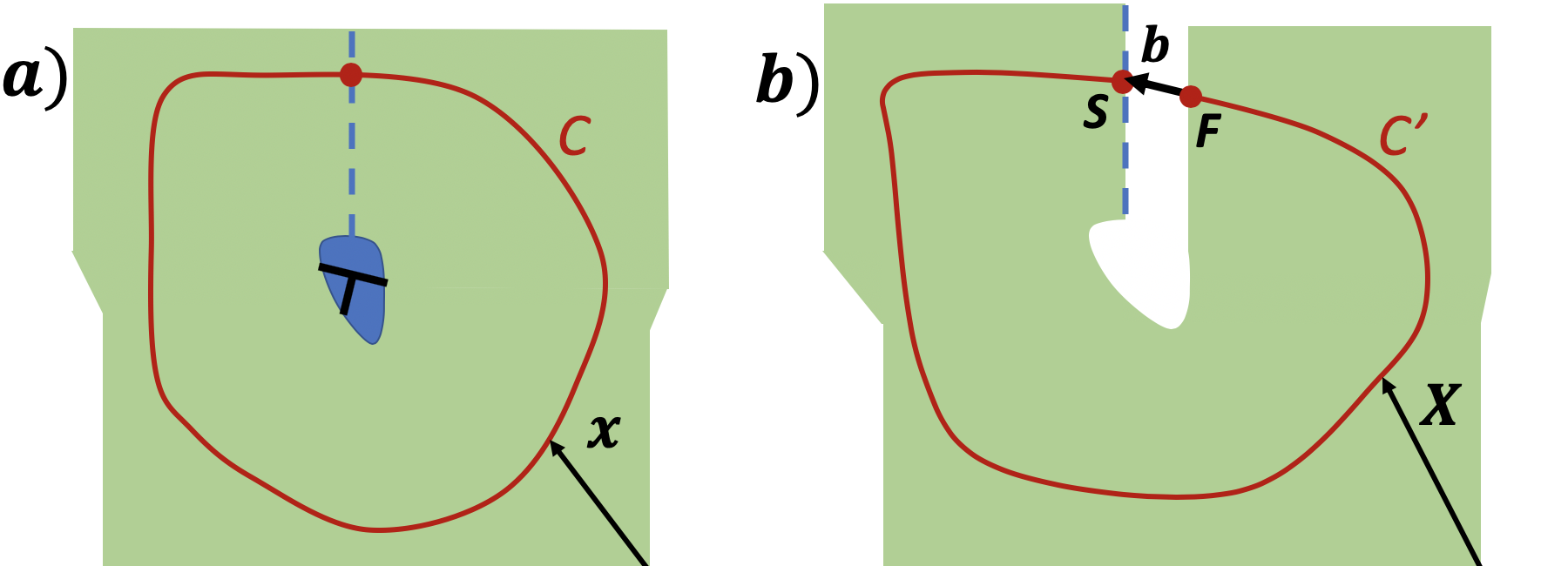}
\caption{Illustration of the determination of true Burgers content for a generic continuum. Panel a depicts the current system in which a core region (blue region) associated with a dislocation is present. In panel b the circuit in the reference system is drawn, where no dislocation is present. $\bm{S}$ denotes the starting position of the circuit, and $\bm{F}$ the final position of the open circuit.} 
\label{fig:burgers_circuit_phase_junction}
\end{figure}

We begin the derivation of this method by defining the Burgers vector, $\bm{b}$, which characterizes the strength of a dislocation. Let us imagine two systems labeled the \textit{reference} and \textit{current} systems, which are characterized by position vectors $\bm{X}$ and $\bm{x}$ respectively. The reference system corresponds to an undeformed system, while the current system corresponds to a deformed, defect containing, system. A position in the reference system can be mapped to a position in the current system by means of the deformation gradient, $\bm{A}$, defined as
\begin{equation}
d x_i = A_{ij} d X_{j}.
\end{equation}
with Einstein summation notation assumed here and for all Latin subscripts throughout this work. Thus, the deformation gradient can be defined as $A_{ij} = \frac{\partial x_i}{\partial X_j}$, and its inverse will be defined as $A^{-1}_{pq} = \frac{\partial X_p}{\partial x_q}$. A dislocation is a line defect in a solid characterized by a discontinuity in the displacement field between the reference and current systems, with the magnitude of this discontinuity being the Burgers vector. The true Burgers vector can be defined by drawing a closed circuit, $C$, around a dislocation as shown in Fig.~\ref{fig:burgers_circuit_phase_junction}. The mathematical expression for this procedure is given as a line integral in the current space \cite{HirthLothe}:
\begin{subequations}\label{eq:def_burgers}
\begin{align}
\bm{b} &= -\oint_{C} \bm{A}^{-1} \cdot d\bm{x},\\
b_k &= - \oint_{C} A^{-1}_{km} dx_m.
\end{align}
\end{subequations}
Equation (\ref{eq:def_burgers}b) is the same as Eq.~(\ref{eq:def_burgers}a), but written in index form.

By applying Stokes' Theorem to Eq.~(\ref{eq:def_burgers}) we find that
\begin{subequations}\label{eq:stokes}
\begin{align}
\bm{b} &= - \int_a \nabla\times \bm{A}^{-1} d\bm{a},\\
b_k &= - \int_a e_{kpq} \frac{\partial A^{-1}_{qm}}{\partial x_p} da_m,
\end{align}
\end{subequations}
where $\bm{e}$ is the Levi-Civita symbol, $a$ is the area encompassed by the circuit $C$, and $d \bm{a} = \hat{\bm{n}} da $, with $\hat{\bm{n}}$ being a unit vector normal to the plane of the differential area, $da$. Equation (\ref{eq:stokes}) makes it clear that the Burgers vector is a flux through an area, and that the density of the Burgers vector is represented by a tensor defined as 
\begin{equation}\label{eq:nye}
\alpha_{km} = -e_{kpq} \frac{\partial A^{-1}_{qm}}{\partial x_p}.
\end{equation}
 
Also known as the Nye tensor, $\alpha_{km}$, gives the Burgers vector component along the $\hat{\bm{e}}_k$ direction of a dislocation with a line direction oriented along the $\hat{\bm{e}}_m$ direction for an area $da$. This means that the diagonal elements of $\bm{\alpha}$ refer to the dislocation density of screw components, while the off-diagonal elements of $\bm{\alpha}$ refer to the dislocation density of edge components. 

Before we evaluate the atomic Nye tensor at the GB, it is important to introduce the concept of the displacement shift complete (DSC) lattice. Assuming that the two lattice vectors that define the two grains which meet at a GB are related by a transformation matrix that is composed of rational numbers, the dichromatic pattern formed by the two lattices will contain a lattice of coincident sites. All the points of the dichromatic pattern comprise of sites of another lattice, called the DSC lattice. The points of the DSC lattice can be described by a linear combination of the two sets of lattice vectors that form the dichromatic pattern associated with the given GB of interest \cite{GRIMMER19741221}:
\begin{equation}\label{eq:dsclattice}
    \bm{R}^{DSC} = m \bm{a}^1 + n\bm{a}^2 + p \bm{a}^3 +  u\bm{q}^1 + v\bm{q}^2 + w\bm{q}^3,
\end{equation}
where $m$, $n$, $p$, $u$, $v$, and $w$ integers, while $\bm{a}^j$ and $\bm{q}^k$ are the sets of lattice vectors associated with the two lattices that form the dichromatic pattern. Using  Eq.~(\ref{eq:dsclattice}) a DSC lattice can be generated and its primitive lattice vectors then determined. In calculating the Nye tensor at a grain boundary we choose the reference system to be the DSC lattice that most closely corresponds to the dichromatic pattern associated with the grain boundary in question. The DSC lattice is a convenient choice for the reference system, because atoms at the GB do not correspond to the lattice of either of the abutting grains, but the GB atoms usually correspond closely to the DSC lattice sites. 

\section{Method}
\label{sec:method}

To calculate $\bm{\alpha}$ for a given atom it is necessary to calculate the deformation gradient. There are numerous methods in place to do this \cite{PhysRevE.57.7192, Stukowski_2012b, Larsen_2016}, but for simplicity we follow the approach outlined by Hartley and Mishin \cite{HARTLEY20051313}. To begin the treatment we consider the $\gamma^{th}$ atom and find all neighbors to this atom within a cutoff radius $R_c$ for the current configuration. For the systems considered here, visualizing Burgers content works well with the settings $R_c=2a_0$ for BCC and $R_c=1.35 a_0$ for FCC, where $a_0$ is the lattice parameter. The set of $n$ neighbor distances within a radius of $R_c$ of the $\gamma^{th}$ atom forms an $n\times 3$ matrix denoted as $\bm{Q}$. We then find the DSC lattice sites relative to $\bm{X}^{(\gamma)}$ approximating each row of $\bm{Q}$ to construct the $n\times 3$ matrix $\bm{P}$. This correspondence is readily determined using the primitive lattice vectors of the DSC lattice.

To ensure stable results, there must be a one-to-one mapping between $\bm{Q}$ and $\bm{P}$. In the case that more than one atom in $\bm{Q}$ corresponds to the same DSC lattice site, the angle $\phi^j$ formed from the $j^{th}$ row of $\bm{Q}$ and $\bm{P}$ are calculated for the degenerate pairs, and the atom for which $\phi$ is smallest is kept in $\bm{Q}$, while all other duplicate entries are deleted from both $\bm{Q}$ and $\bm{P}$. The DSC lattice is fine enough that this rarely happens. In addition, if any neighboring atom has a Q-P angle greater than $\phi_{max}=27^{\circ}$ this atom is not considered as a neighbor, and the row is deleted from $\bm{Q}$ and $\bm{P}$. The inverse of the deformation gradient for the $\gamma^{th}$ atom, shown in Eq.~(\ref{eq:nye}), relates $\bm{Q}$ and $\bm{P}$ via:
\begin{equation}\label{eq:pseudo}
\bm{P} = \bm{Q} \cdot \bm{A}^{-T}(\gamma),
\end{equation} 
\noindent with $\bm{A}^{-T}$ being the transpose of $\bm{A}^{-1}$.  $\bm{A}^{-1}(\gamma)$ is determined by applying the pseudo-inverse of $\bm{Q}$ to both sides of Eq.~(\ref{eq:pseudo}) and then taking the transpose:
\begin{equation}\label{eq:defgradformula}
\bm{A}^{-1}(\gamma) = (\bm{Q}^{+} \cdot \bm{P})^T.
\end{equation} 
The pseudo-inverse (Moore-Penrose inverse) is calculated using the following formula:
\begin{equation}\label{eq:pseudocalc}
\bm{Q}^{+} = \left( \bm{Q}^T \cdot \bm{Q} \right)^{-1} \cdot \bm{Q}^T .
\end{equation} 
Here $\left( \bm{Q}^T \cdot \bm{Q} \right)^{-1}$ is the usual $3\times 3$ matrix inverse.

To find $\bm{\alpha}(\gamma)$ it is necessary to take the curl of $\bm{A}^{-1}(\gamma)$. Hartley and Mishin achieve this by first creating an $n\times 3\times 3$ array termed $\Delta A^{-1}_{\beta jk} = A^{-1}_{jk}(\beta) - A^{-1}_{jk}(\gamma)$. They then show that Eq.~(\ref{eq:nye}) can be written as 
\begin{equation}
\alpha_{km}(\gamma) = \sum_{\beta=1}^n e_{kpq} Q^{+}_{p\beta} \Delta A^{-1}_{\beta q m}.
\end{equation}

\begin{center}
\begin{figure}[ht!]
\centering
\includegraphics[width=\textwidth]{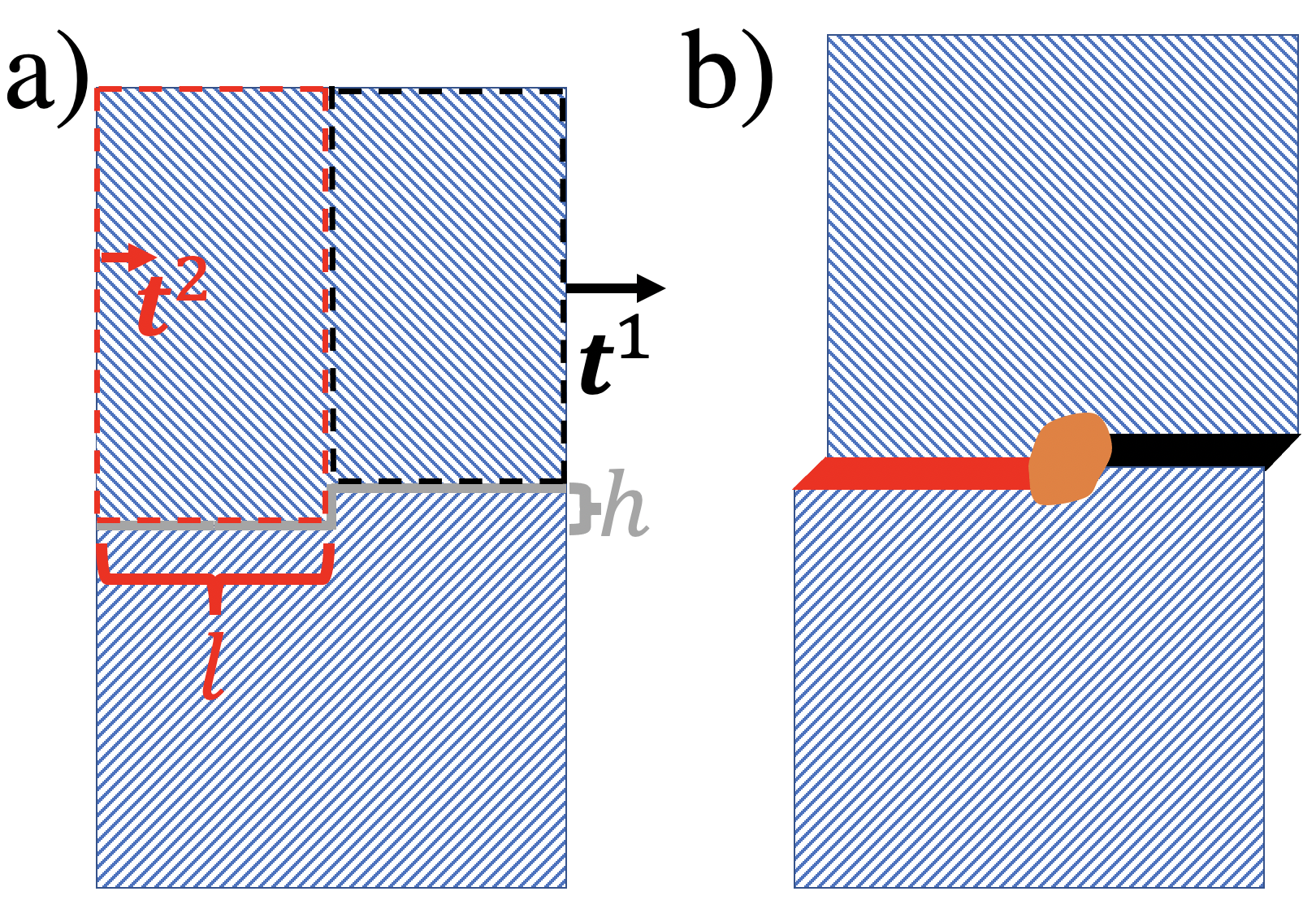}
\caption{Illustration of the setup of the molecular dynamics simulations for the 1D case. Panel a depicts two crystals, separated by an interface with a step of height $h$, being displaced relative to each other. By displacing a section of the upper crystal of length $l$ by $\bm{t}^2$ and the rest of the upper crystal by the vector $\bm{t}^1$. Panel b depicts a dual phase grain boundary created after relaxing the system. The core of the grain boundary junction is signified by the orange region. }
\label{fig:md_simulation_setup}
\end{figure}
\end{center}

As examples, we produced disconnections, be they grain boundary phase junctions or not, using the procedure of Winter et al.\ \cite{winter2021nucleation} by first generating a bicrystal with a step of height $h$ along the grain boundary. The top crystal was then displaced relative to the lower crystal by $\bm{t}^1 $, as shown in Fig.~\ref{fig:md_simulation_setup}a. Then a region of the upper crystal, of length $l$ along the $\hat{\bm{e}}_1$-direction, was further displaced by $\bm{t}^2$. The combined system is then relaxed using LAMMPS \cite{THOMPSON2022108171} such that the maximum force between atoms was $10^{-3}$ eV$/$\AA, with the resulting final system illustrated in Fig.~\ref{fig:md_simulation_setup}b.

\section{Results and Discussion}
\label{sec:results}


To illustrate the effectiveness of using the Nye tensor to detect dislocation content within a grain boundary we present two examples in BCC tungsten and one example in FCC copper modeled with the embedded atom method potential developed by Zhou et al.\ \cite{ZHOU20014005} and Mishin et al., respectively \cite{PhysRevB.63.224106}. In the case of the W, the two examples consist of disconnections in a $\Sigma 5(210)[001]$ tilt boundary and a grain boundary phase junction for a $\Sigma 17 (410)[001]$ tilt boundary. In the case of the grain boundary phase junction we will consider both a 1D example, as shown in Fig.~\ref{fig:md_simulation_setup} and a fully 2D grain boundary phase nucleus. For the case of Cu, we will consider a $\Sigma 5(210)[001]$ GB phase junction consisting of a normal kite  and a split kite structure \cite{Frolov2013}. The values for the relevant parameters, described in Section \ref{sec:method}, used to construct the two examples are listed in Table \ref{tab:parameters}. In Table \ref{tab:parameters} the terms $f^1$ and $f^2$ correspond to the fraction of atoms deleted from the atomic plane at the grain boundary. Unlike in the case of W, in which relative displacements are used to create disconnection content, the Cu GB phase junction is created entirely by deleting atoms at the grain boundary to create the split kite structure.

\begin{table}
\caption{\label{tab:parameters} Parameters used to construct the example structures. }
\begin{tabular}{cccccccc}
GB & $h\ (\mathrm{\AA})$ & $t^1_1\ (\mathrm{\AA})$  & $t^1_2\ (\mathrm{\AA})$ & $t^2_1\ (\mathrm{\AA})$  & $t^2_2\ (\mathrm{\AA})$ & $f^1$ & $f^2$\\
\hline
W $\Sigma 5(210)[001]$  & 1.583 & 0.000 & 1.265 &0.000 & 0.633 & 0.000 & 0.000\\
W $\Sigma 17(410)[001]$  & 0.000 & 1.957 & 0.316 & 0.652 & 0.000 & 0.000 & 0.000\\
Cu $\Sigma 5 (210)[001]$ & 0.000 & 0.000 & 0.000 & 0.000 & 0.000 & 0.000 & 0.5000
\end{tabular}
\end{table}

\begin{center}
\begin{figure}[ht!]
\centering
\includegraphics[width=\textwidth]{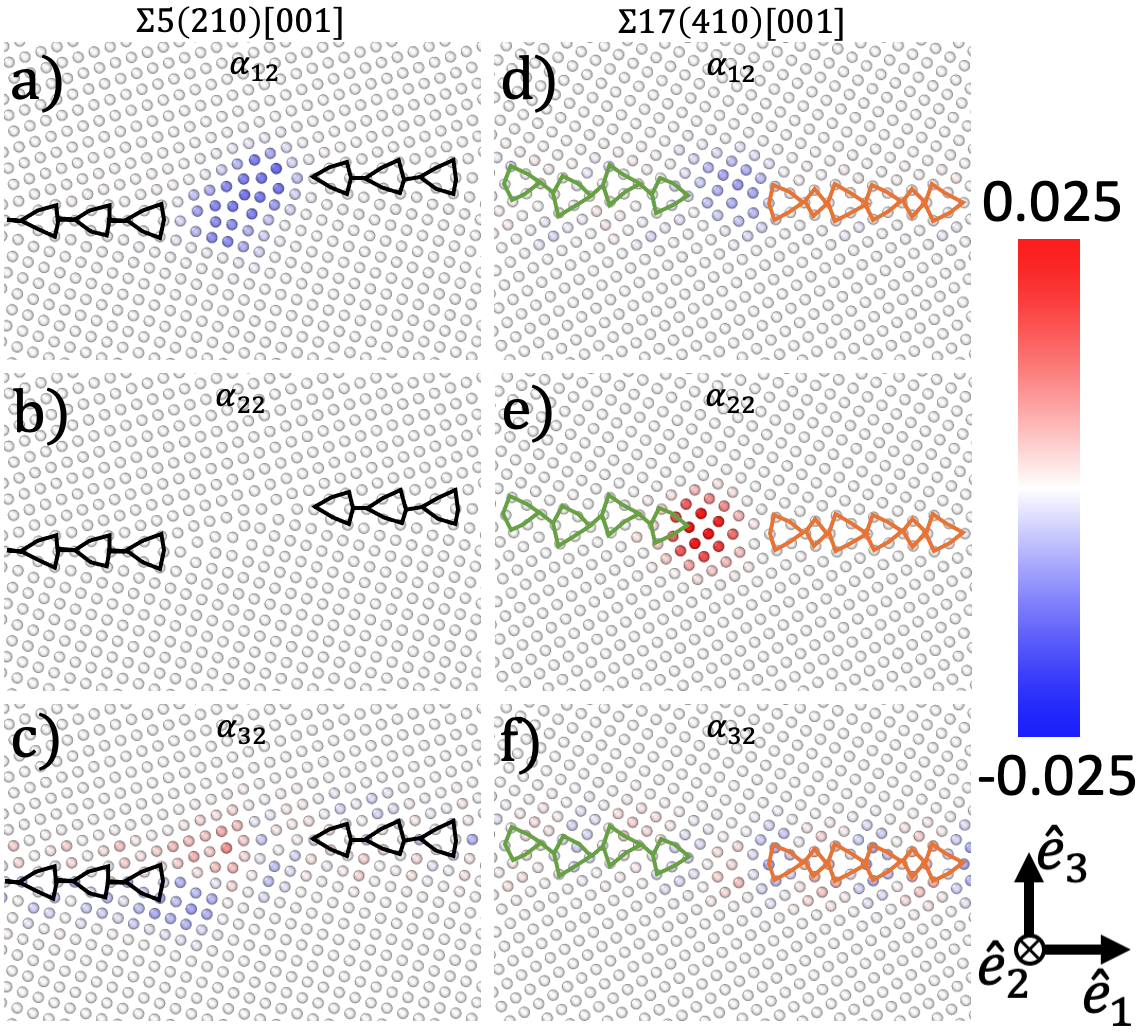}
\caption{Calculation of the Nye tensor at each atom for a W $\Sigma5(210)[001]$ disconnection, panels a-c, and W $\Sigma17(410)[001]$ grain boundary phase junction, panels d-f. The guidelines are shown to illustrate the structure of the grain boundary phase. The units of Nye tensor are \AA$^{-1}$.}
\label{fig:nye_tensor_2d}
\end{figure}
\end{center}

The results for the atomic Nye tensor of a disconnection in a $\Sigma 5(210)[001]$ GB in Fig.~\ref{fig:nye_tensor_2d}a-c (left column) are very good. The Nye tensor values in the disconnection are much larger than the noise in the grain boundary and bulk, providing a clear visualization. To test whether the values correspond to Burgers content, we use the analysis of Frolov et al.\ \cite{FrolovBurgers} to calculate the Burgers vector of the disconnection, and find it to be $\bm{b} = -\frac{\sqrt{5}a_0}{5}\ \hat{\bm{e}}_1$, with $a_0 = 3.165$\ \AA. Qualitatively, the atomic Nye tensor analysis agrees with this assessment. From Fig.~\ref{fig:nye_tensor_2d}a there appears to be a high density of $b_1$, and Fig.~\ref{fig:nye_tensor_2d}b shows that there is no $b_2$. While there appears to be a noticeable density of $b_3$ in \ref{fig:nye_tensor_2d}c, it is relatively small compared to the density of $b_1$ from Fig.~\ref{fig:nye_tensor_2d}. The distribution of $b_3$ appears to be in the form of a dipole, so that if one were to integrate $\alpha_{32}$ over a large enough area, $b_3\approx 0$, in agreement with the Burgers circuit analysis. To illustrate this fact, we numerically integrate the the components of the Nye tensor within a 20 \AA $\times$ 20 \AA\ region centered at the dislocation core to get the resulting Burgers vector. The Burgers vector is found to be $b_1 = -1.35$\ \AA, $b_2 = 0.00$\ \AA, and $b_3 = 0.00$ \AA. This is in excellent agreement with the Burgers circuit analysis, and indeed shows that in the case of Fig.~\ref{fig:nye_tensor_2d}c, the dipoles do cancel out.

For the second example, we consider the atomic Nye tensor for a grain boundary phase junction of a $\Sigma 17 (410)[001]$ tilt boundary. The lower and higher energy grain boundary phases are illustrated in Fig.~\ref{fig:nye_tensor_2d}d-f (right column) as the orange and green phases respectively. Again, the contrast of the disconnection against the grain boundary and bulk is high. A Burgers circuit analysis of this grain boundary gives a value of $\bm{b} = -0.753\ \mathrm{\AA} \hat{\bm{e}}_1 + 1.582\ \mathrm{\AA} \hat{\bm{e}}_2 + 0.012\ \mathrm{\AA} \hat{\bm{e}}_3$. In agreement, the atomic Nye tensor plots show significant Burgers content related to $b_1$ and $b_2$, but little related to $b_3$.  Again, performing integration of the three Nye tensor components over a 20 \AA $\times$ 20 \AA\ area centered at the GB phase junctions gives $b_1 = -0.57$\ \AA, $b_2 = 1.62$ \AA\ and $b_3 = 0.00$ \AA, which is in good agreement with the Burgers circuit analysis.

\begin{center}
\begin{figure}[ht!]
\centering
\includegraphics[width=\textwidth]{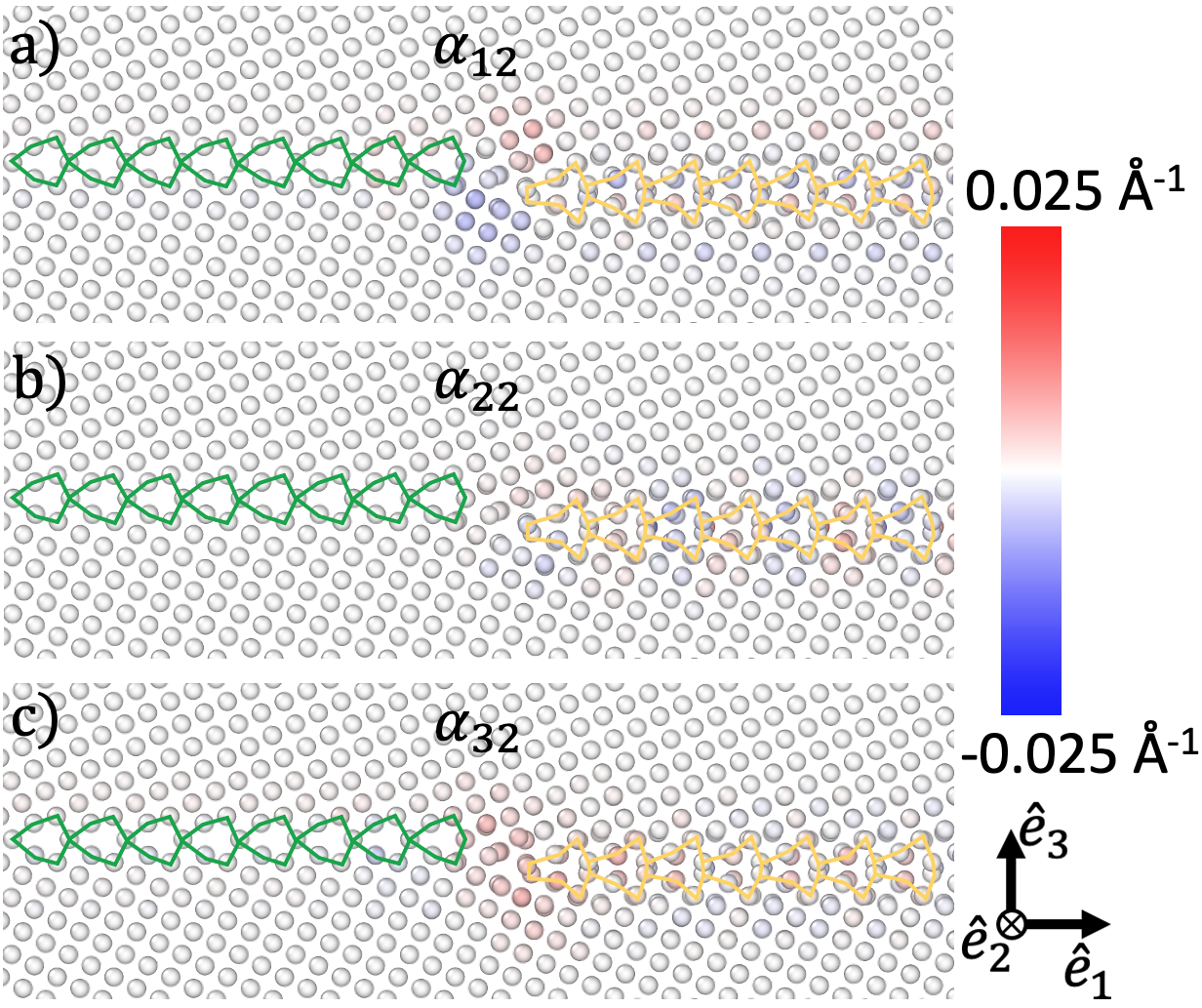}
\caption{Calculation of the Nye tensor at each atom for a $\Sigma5(210)[001]$ GB phase junction in FCC Cu, panels a-c The guidelines are shown to illustrate the structure of the grain boundary phase. The units of Nye tensor are \AA$^{-1}$.}
\label{fig:nye_tensor_cu}
\end{figure}
\end{center}

For the third example, we consider the atomic Nye tensor for a grain boundary phase junction of a $\Sigma 5 (210)[001]$ symmetric tilt boundary in FCC Cu. This system is an important test, because it involves the split kite structure, which is structurally more complicated than the other GB structures considered in this work. The results of the analysis are shown in Fig.~\ref{fig:nye_tensor_cu}. A Burgers circuit analysis of this structure gives the Burgers vector of the GB phase junction to be $\bm{b} = 0.22\ \mathrm{\AA} \hat{\bm{e}}_1 + 0.57\ \mathrm{\AA} \hat{\bm{e}}_3$. From Fig.~\ref{fig:nye_tensor_cu}b it is apparent that there is no screw Burgers content. Figures \ref{fig:nye_tensor_cu}a and \ref{fig:nye_tensor_cu}c appear to show edge content for both the $b_1$ and $b_3$ components. The Burgers content of $b_3$ at the GB phase junction is especially evident. It is less evident for the $b_1$ component in Fig.~\ref{fig:nye_tensor_cu}a. This is likely due to the smaller Burgers content measured from the Burgers circuit analysis ($b_1 = 0.22$ \AA). Performing the  numerical integration procedure over the Nye tensor components gives $b_1 = -0.01$\AA, $b_2 = 0.00$\AA, and $b_3 = 0.16$\AA. Unlike the prior BCC cases, the estimate of the Burgers content from the atomic Nye tensor is not quantitatively correct in this case. The results, however, appear to be qualitatively correct as the $b_3$ component has a much larger magnitude than the other two components.

\begin{center}
\begin{figure}[ht!]
\centering
\includegraphics[width=\textwidth]{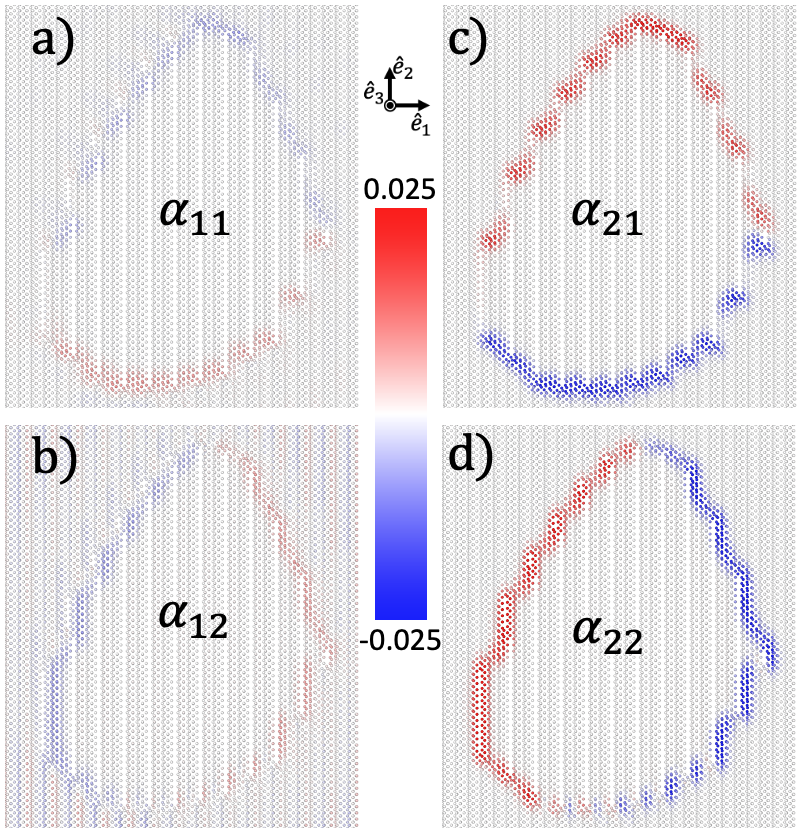}
\caption{Calculation of the Nye tensor at each atom for a W $\Sigma17(410)[001]$ grain boundary phase junction. Each panel illustrates a different component of the atomic Nye tensor. Panels a and c refer to the Burgers vector density for a dislocation with a line direction of $\xi=\hat{\bm{e}}_1$. Panels b and d refer to the Burgers vector density for a dislocation with a line direction of $\xi=\hat{\bm{e}}_2$. The units of Nye tensor are \AA$^{-1}$.}
\label{fig:nye_circular}
\end{figure}
\end{center}

As a final example we consider a fully 2D grain boundary phase nucleus as shown in Fig.~\ref{fig:nye_circular}. The nucleus was constructed using the methodology of Winter et al.\ \cite{winter2021nucleation}, with the initial radius of the stable grain boundary phase nucleus being approximately 131 \AA. Note that the frame has changed from Fig.~\ref{fig:nye_tensor_2d}, and now the line direction is in the plane of the figure. Again there is good contrast in Fig.~\ref{fig:nye_circular} of the phase junction against the surrounding material. It is clear the nucleus relaxes away from its initial circular shape, minimizing the energy including elastic and interfacial contributions \cite{winter2021nucleation}. It is also clear that the curvature of the phase junction bounding the nucleus is brought about by the formation of kinks, in agreement with the model of interface migration proposed by Han et al.\ \cite{han2021disconnectionmediated}. As Fig.~\ref{fig:nye_circular}c illustrates, the atomic Nye tensor proves to be an excellent visualization tool for the kinked disconnection. The visualization still works because the segments of the kinked phase junction contain the same dislocation content as the straight junction shown in Fig.~\ref{fig:nye_tensor_2d}d-f, but a different dislocation line direction, $\hat{\bm{e}}_2$ instead of $\hat{\bm{e}}_1$.

\section{Discussion and Conclusion}
\label{sec:conclusion}

Here we have introduced a powerful new tool for visualizing dislocation content at grain boundaries. The atomic Nye tensor allows us to elucidate a dislocation's character, as well as its relative strength. This will enable more in-depth studies of the basic mechanisms governing microstructural evolution. In considering a dual grain system, the lattice sites of the two grains sit on the DSC lattice. This is significant as it means that the atomic Nye tensor, as described here, can equally describe disconnections and bulk dislocations (and as a result low-angle GBs), meaning that the atomic Nye tensor is an ideal tool to use in simulations of the interaction between bulk dislocations and grain boundaries. 

As noted above, there are some limitations to the use of the atomic Nye tensor to visualize disconnections, most clearly seen in the formation of local dipoles in Fig.~\ref{fig:nye_tensor_2d}c and Fig.~\ref{fig:nye_tensor_cu}. The origin of these local dipoles is likely due to the fact that the atoms within the GB stray from the DSC lattice. The existing DXA method can add additional clarity in some of these cases. We note that this work is meant as a proof of concept, and there are likely to be approaches for improving the method. For instance, a range of methods has been developed to accurately calculate the atomic strain, and as a result, deformation gradient \cite{PhysRevE.57.7192, Stukowski_2012b, Larsen_2016}. We have found that increasing the neighbor distance used in assessing the atomic deformation gradient is an efficient approach for smoothing out the local dipoles at the grain boundary, but is not necessarily the ideal approach. However, we leave the task of optimizing this approach to future work.

Another way the atomic Nye tensor could be implemented to enable these studies is in an extension to DXA to facilitate its application to disconnections in grain boundaries. Methods such as Polyhedral Template Matching \cite{Larsen_2016} (or earlier techniques \cite{rudd2010grainorient}) allow for the relative orientations of grains to be calculated. This in turn allows for the DSC lattice of a given grain boundary to be determined \cite{GRIMMER19741221}, and thus leads to the atomic Nye tensor, as described in this work. The atomic Nye tensor can then be used to determine the ``bad" regions that DXA requires to perform its Burgers circuit analysis. Such an implementation remains for future work.

The method we have introduced to visualize dislocations within grain boundaries applies the concept of the displacement shift complete lattice to the atomic Nye tensor. As a result, regions of a grain boundary that do not contain dislocation content, but also are not perfectly crystalline, do not register as being within the core of a defect. A great advantage of this method is that it requires no training to recognize disconnections, as it relies only on the macroscopic degrees of freedom of the grain boundary to generate the DSC lattice. This method has been shown here to be effective in visualizing a disconnection along a $\Sigma 5(210)[001]$ tilt grain boundary, and a grain boundary phase junction in a $\Sigma 17(410)[001]$ tilt grain boundary, both in BCC W. The method has also been applied here to look at a $\Sigma 5(210)[001]$ phase junction in Cu. This method used with other methods, such as the dislocation extraction algorithm and polyhedral template matching, promises to be an effective tool in analyzing large-scale atomistic simulations of microstructural evolution and plasticity.

\section{Acknowledgments}
This work was performed under the auspices of the U.S. Department of Energy by Lawrence Livermore National Laboratory under Contract DE-AC52-07NA27344. This material is based upon work supported by the U.S. Department of Energy, Office of Science, Office of Fusion Energy Sciences. Computing support for this work came from the Lawrence Livermore National Laboratory Institutional Computing Grand Challenge program.

 \bibliographystyle{elsarticle-num} 
 \bibliography{cas-refs_rer}

\providecommand{\noopsort}[1]{}\providecommand{\singleletter}[1]{#1}%
\begin{thebibliography}{10}
\expandafter\ifx\csname url\endcsname\relax
  \def\url#1{\texttt{#1}}\fi
\expandafter\ifx\csname urlprefix\endcsname\relax\def\urlprefix{URL }\fi
\expandafter\ifx\csname href\endcsname\relax
  \def\href#1#2{#2} \def\path#1{#1}\fi

\bibitem{Shober1970}
T.~Schober, R.~W. Balluffi, Quantitative observation of misfit dislocation
  arrays in low and high angle twist grain boundaries, The Philosophical
  Magazine: A Journal of Theoretical Experimental and Applied Physics 21~(169)
  (1970) 109--123.
\newblock \href {https://doi.org/10.1080/14786437008238400}
  {\path{doi:10.1080/14786437008238400}}.

\bibitem{HIRTH1973929}
J.~Hirth, R.~Balluffi, On grain boundary dislocations and ledges, Acta
  Metallurgica 21~(7) (1973) 929--942.
\newblock \href {https://doi.org/10.1016/0001-6160(73)90150-8}
  {\path{doi:10.1016/0001-6160(73)90150-8}}.

\bibitem{Sun1982}
C.~P. Sun, R.~W. Balluffi, Secondary grain boundary dislocations in [001] twist
  boundaries in {MgO} {I.} {Intrinsic} structures, Philosophical Magazine A
  46~(1) (1982) 49--62.
\newblock \href {https://doi.org/10.1080/01418618208236206}
  {\path{doi:10.1080/01418618208236206}}.

\bibitem{BABCOCK19892357}
S.~Babcock, R.~Balluffi, Grain boundary kinetics—{I.} {In} situ observations
  of coupled grain boundary dislocation motion, crystal translation and
  boundary displacement, Acta Metallurgica 37~(9) (1989) 2357--2365.
\newblock \href {https://doi.org/10.1016/0001-6160(89)90033-3}
  {\path{doi:10.1016/0001-6160(89)90033-3}}.

\bibitem{BABCOCK19892367}
S.~Babcock, R.~Balluffi, Grain boundary kinetics—{II.} {In} situ observations
  of the role of grain boundary dislocations in high-angle boundary migration,
  Acta Metallurgica 37~(9) (1989) 2367--2376.
\newblock \href {https://doi.org/10.1016/0001-6160(89)90034-5}
  {\path{doi:10.1016/0001-6160(89)90034-5}}.

\bibitem{HIRTH19964749}
J.~Hirth, R.~Pond, Steps, dislocations and disconnections as interface defects
  relating to structure and phase transformations, Acta Materialia 44~(12)
  (1996) 4749--4763.
\newblock \href {https://doi.org/10.1016/S1359-6454(96)00132-2}
  {\path{doi:10.1016/S1359-6454(96)00132-2}}.

\bibitem{HAN2018386}
J.~Han, S.~L. Thomas, D.~J. Srolovitz, Grain-boundary kinetics: {A} unified
  approach, Progress in Materials Science 98 (2018) 386--476.
\newblock \href {https://doi.org/10.1016/j.pmatsci.2018.05.004}
  {\path{doi:10.1016/j.pmatsci.2018.05.004}}.

\bibitem{Chen4533}
K.~Chen, J.~Han, X.~Pan, D.~J. Srolovitz, The grain boundary mobility tensor,
  Proceedings of the National Academy of Sciences 117~(9) (2020) 4533--4538.
\newblock \href {https://doi.org/10.1073/pnas.1920504117}
  {\path{doi:10.1073/pnas.1920504117}}.

\bibitem{CHEN2020412}
K.~Chen, J.~Han, D.~J. Srolovitz, On the temperature dependence of grain
  boundary mobility, Acta Materialia 194 (2020) 412--421.
\newblock \href {https://doi.org/10.1016/j.actamat.2020.04.057}
  {\path{doi:10.1016/j.actamat.2020.04.057}}.

\bibitem{Thomas8756}
S.~L. Thomas, C.~Wei, J.~Han, Y.~Xiang, D.~J. Srolovitz, Disconnection
  description of triple-junction motion, Proceedings of the National Academy of
  Sciences 116~(18) (2019) 8756--8765.
\newblock \href {https://doi.org/10.1073/pnas.1820789116}
  {\path{doi:10.1073/pnas.1820789116}}.

\bibitem{winter2021nucleation}
I.~S. Winter, R.~E. Rudd, T.~Oppelstrup, T.~Frolov, Nucleation of grain
  boundary phases, Phys. Rev. Lett. 128 (2022) 035701.
\newblock \href {https://doi.org/10.1103/PhysRevLett.128.035701}
  {\path{doi:10.1103/PhysRevLett.128.035701}}.

\bibitem{Chen33077}
K.~Chen, D.~J. Srolovitz, J.~Han, Grain-boundary topological phase transitions,
  Proceedings of the National Academy of Sciences 117~(52) (2020) 33077--33083.
\newblock \href {https://doi.org/10.1073/pnas.2017390117}
  {\path{doi:10.1073/pnas.2017390117}}.

\bibitem{Zepeda2017}
L.~A. Zepeda-Ruiz, A.~Stukowski, T.~Oppelstrup, V.~V. Bulatov, Probing the
  limits of metal plasticity with molecular dynamics simulations, Nature
  550~(7677) (2017) 492--495.
\newblock \href {https://doi.org/10.1038/nature23472}
  {\path{doi:10.1038/nature23472}}.

\bibitem{wehrenberg2017grainrot}
C.~E. Wehrenberg, D.~McGonegle, C.~Bolme, A.~Higginbotham, A.~Lazicki, H.~J.
  Lee, B.~Nagler, H.-S. Park, B.~A. Remington, R.~E. Rudd, M.~Sliwa, M.~Suggit,
  D.~Swift, F.~Tavella, L.~Zepeda-Ruiz, J.~S. Wark, In situ x-ray diffraction
  measurement of shock-wave-driven twinning and lattice dynamics, Nature 550
  (2017) 496--499.
\newblock \href {https://doi.org/10.1038/nature24061}
  {\path{doi:10.1038/nature24061}}.

\bibitem{Zepeda2021}
L.~A. Zepeda-Ruiz, A.~Stukowski, T.~Oppelstrup, N.~Bertin, N.~R. Barton,
  R.~Freitas, V.~V. Bulatov, Atomistic insights into metal hardening, Nature
  Materials 20~(3) (2021) 315--320.
\newblock \href {https://doi.org/10.1038/s41563-020-00815-1}
  {\path{doi:10.1038/s41563-020-00815-1}}.

\bibitem{Stukowski_2010}
A.~Stukowski, K.~Albe, Extracting dislocations and non-dislocation crystal
  defects from atomistic simulation data, Modelling and Simulation in Materials
  Science and Engineering 18~(8) (2010) 085001.
\newblock \href {https://doi.org/10.1088/0965-0393/18/8/085001}
  {\path{doi:10.1088/0965-0393/18/8/085001}}.

\bibitem{Stukowski_2012}
A.~Stukowski, V.~V. Bulatov, A.~Arsenlis, Automated identification and indexing
  of dislocations in crystal interfaces, Modelling and Simulation in Materials
  Science and Engineering 20~(8) (2012) 085007.
\newblock \href {https://doi.org/10.1088/0965-0393/20/8/085007}
  {\path{doi:10.1088/0965-0393/20/8/085007}}.

\bibitem{NYE1953153}
J.~Nye, Some geometrical relations in dislocated crystals, Acta Metallurgica
  1~(2) (1953) 153--162.
\newblock \href {https://doi.org/10.1016/0001-6160(53)90054-6}
  {\path{doi:10.1016/0001-6160(53)90054-6}}.

\bibitem{HARTLEY20051313}
C.~Hartley, Y.~Mishin, Characterization and visualization of the lattice misfit
  associated with dislocation cores, Acta Materialia 53~(5) (2005) 1313--1321.
\newblock \href {https://doi.org/10.1016/j.actamat.2004.11.027}
  {\path{doi:10.1016/j.actamat.2004.11.027}}.

\bibitem{HARTLEY200518}
C.~S. Hartley, Y.~Mishin, Representation of dislocation cores using nye tensor
  distributions, Materials Science and Engineering: A 400-401 (2005) 18--21,
  dislocations 2004.
\newblock \href {https://doi.org/10.1016/j.msea.2005.03.076}
  {\path{doi:10.1016/j.msea.2005.03.076}}.

\bibitem{HirthLothe}
J.~Hirth, J.~Lothe, Theory of {D}islocations, 2nd Edition, Krieger, Malabar,
  Florida, 1982, Ch.~5, p. 117.

\bibitem{GRIMMER19741221}
H.~Grimmer, A reciprocity relation between the coincidence site lattice and the
  {DSC} lattice, Scripta Metallurgica 8~(11) (1974) 1221 -- 1223.
\newblock \href {https://doi.org/10.1016/0036-9748(74)90334-2}
  {\path{doi:10.1016/0036-9748(74)90334-2}}.

\bibitem{PhysRevE.57.7192}
M.~L. Falk, J.~S. Langer, Dynamics of viscoplastic deformation in amorphous
  solids, Phys. Rev. E 57 (1998) 7192--7205.
\newblock \href {https://doi.org/10.1103/PhysRevE.57.7192}
  {\path{doi:10.1103/PhysRevE.57.7192}}.

\bibitem{Stukowski_2012b}
A.~Stukowski, A.~Arsenlis, On the elastic{\textendash}plastic decomposition of
  crystal deformation at the atomic scale, Modelling and Simulation in
  Materials Science and Engineering 20~(3) (2012) 035012.
\newblock \href {https://doi.org/10.1088/0965-0393/20/3/035012}
  {\path{doi:10.1088/0965-0393/20/3/035012}}.

\bibitem{Larsen_2016}
P.~M. Larsen, S.~Schmidt, J.~Schi{\o}tz, Robust structural identification via
  polyhedral template matching, Modelling and Simulation in Materials Science
  and Engineering 24~(5) (2016) 055007.
\newblock \href {https://doi.org/10.1088/0965-0393/24/5/055007}
  {\path{doi:10.1088/0965-0393/24/5/055007}}.

\bibitem{THOMPSON2022108171}
A.~P. Thompson, H.~M. Aktulga, R.~Berger, D.~S. Bolintineanu, W.~M. Brown,
  P.~S. Crozier, P.~J. {in 't Veld}, A.~Kohlmeyer, S.~G. Moore, T.~D. Nguyen,
  R.~Shan, M.~J. Stevens, J.~Tranchida, C.~Trott, S.~J. Plimpton, Lammps - a
  flexible simulation tool for particle-based materials modeling at the atomic,
  meso, and continuum scales, Computer Physics Communications 271 (2022)
  108171.
\newblock \href {https://doi.org/10.1016/j.cpc.2021.108171}
  {\path{doi:10.1016/j.cpc.2021.108171}}.

\bibitem{ZHOU20014005}
X.~Zhou, H.~Wadley, R.~Johnson, D.~Larson, N.~Tabat, A.~Cerezo,
  A.~Petford-Long, G.~Smith, P.~Clifton, R.~Martens, T.~Kelly, Atomic scale
  structure of sputtered metal multilayers, Acta Materialia 49~(19) (2001) 4005
  -- 4015.
\newblock \href {https://doi.org/10.1016/S1359-6454(01)00287-7}
  {\path{doi:10.1016/S1359-6454(01)00287-7}}.

\bibitem{PhysRevB.63.224106}
Y.~Mishin, M.~J. Mehl, D.~A. Papaconstantopoulos, A.~F. Voter, J.~D. Kress,
  Structural stability and lattice defects in copper: Ab initio, tight-binding,
  and embedded-atom calculations, Phys. Rev. B 63 (2001) 224106.
\newblock \href {https://doi.org/10.1103/PhysRevB.63.224106}
  {\path{doi:10.1103/PhysRevB.63.224106}}.

\bibitem{Frolov2013}
T.~Frolov, D.~L. Olmsted, M.~Asta, Y.~Mishin, {Structural phase transformations
  in metallic grain boundaries}, Nature Communications 4~(1) (2013) 1899.
\newblock \href {https://doi.org/10.1038/ncomms2919}
  {\path{doi:10.1038/ncomms2919}}.

\bibitem{FrolovBurgers}
T.~Frolov, D.~L. Medlin, M.~Asta, Dislocation content of grain boundary phase
  junctions and its relation to grain boundary excess properties, Phys. Rev. B
  103 (2021) 184108.
\newblock \href {https://doi.org/10.1103/PhysRevB.103.184108}
  {\path{doi:10.1103/PhysRevB.103.184108}}.

\bibitem{han2021disconnectionmediated}
J.~Han, D.~J. Srolovitz, M.~Salvalaglio, Disconnection-mediated migration of
  interfaces in microstructures: {I.} continuum model (2021).
\newblock \href {http://arxiv.org/abs/2103.09688} {\path{arXiv:2103.09688}}.

\bibitem{rudd2010grainorient}
R.~E. Rudd, High-rate plastic deformation of nanocrystalline tantalum to large
  strains: Molecular dynamics simulation, Mater. Sci. Forum 633-634 (2010)
  3--19.
\newblock \href {https://doi.org/10.4028/www.scientific.net/MSF.633-634.3}
  {\path{doi:10.4028/www.scientific.net/MSF.633-634.3}}.

\end{thebibliography}





\end{document}